\documentclass[12pt]{article}
\newlength{\mytopmargin}
\newlength{\myleftmargin}
\newtheorem{theorem}{Theorem}[section]
\newtheorem{proposition}[theorem]{Proposition}

\usepackage{amsmath,amsfonts,amssymb}

\usepackage{epsf}
\usepackage{eepic}

\newcommand\psymmU{%
\begin{picture}(1,1)(0,0)%
\allinethickness{0.5pt}%
\path(0,0)(0,1)(1,1)(1,0)(0,0)%
\end{picture}}
\newcommand\psymmUU{%
\begin{picture}(1,1)(0,0)%
\allinethickness{0.5pt}%
\path(0,0)(0,1)(1,1)(1,0)(0,0)%
\put(0.5,0.5){\makebox(0,0){$\cdot$}}%
\end{picture}}
\newcommand\psymmO{%
\begin{picture}(1,1)(0,0)%
\allinethickness{0.5pt}%
\path(0,0)(0,1)(1,1)(1,0)(0,0)%
\path(0,0)(1,1)%
\end{picture}}
\newcommand\psymmS{%
\begin{picture}(1,1)(0,0)%
\allinethickness{0.5pt}%
\path(0,0)(0,1)(1,1)(1,0)(0,0)%
\path(1,0)(0,1)%
\end{picture}}
\newcommand\psymmu{%
\begin{picture}(1,1)(0,0)%
\allinethickness{0.5pt}%
\path(0,0)(0,1)(1,1)(1,0)(0,0)%
\path(0,0)(1,1)%
\path(0,1)(1,0)%
\end{picture}}

\newbox\tsymmUbox
\newbox\tsymmUUbox
\newbox\tsymmObox
\newbox\tsymmSbox
\newbox\tsymmubox
\setbox\tsymmUbox =\hbox{\kern0.75pt\setlength{\unitlength}{6pt}\psymmU \kern0.75pt}

\setbox\tsymmUUbox=\hbox{\kern0.75pt\setlength{\unitlength}{6pt}\psymmUU\kern0.75pt}
\setbox\tsymmObox =\hbox{\kern0.75pt\setlength{\unitlength}{6pt}\psymmO \kern0.75pt}
\setbox\tsymmSbox =\hbox{\kern0.75pt\setlength{\unitlength}{6pt}\psymmS \kern0.75pt}
\setbox\tsymmubox =\hbox{\kern0.75pt\setlength{\unitlength}{6pt}\psymmu \kern0.75pt}

\newbox\symmUbox
\newbox\symmUUbox
\newbox\symmObox
\newbox\symmSbox
\newbox\symmubox
\setbox\symmUbox =\hbox{\kern0.75pt\setlength{\unitlength}{4.5pt}\psymmU \kern0.75pt}
\setbox\symmUUbox=\hbox{\kern0.75pt\setlength{\unitlength}{4.5pt}\psymmUU\kern0.75pt}
\setbox\symmObox =\hbox{\kern0.75pt\setlength{\unitlength}{4.5pt}\psymmO \kern0.75pt}
\setbox\symmSbox =\hbox{\kern0.75pt\setlength{\unitlength}{4.5pt}\psymmS \kern0.75pt}
\setbox\symmubox =\hbox{\kern0.75pt\setlength{\unitlength}{4.5pt}\psymmu \kern0.75pt}

\def\symmO{{\copy\symmObox}}

\begin{document}
\title{Probability densities and distributions for spiked and general variance Wishart $\beta$-ensembles}
\author{Peter J. Forrester}
\date{}
\maketitle

\noindent
\thanks{\small
Department of Mathematics and Statistics,
The University of Melbourne,
Victoria 3010, Australia email: { P.Forrester@ms.unimelb.edu.au} 
}

\begin{abstract}
\noindent A Wishart matrix is said to be spiked when the underlying covariance matrix has a
single eigenvalue $b$ different from unity. As $b$ increases through $b=2$, a gap forms from the largest eigenvalue to the rest of the spectrum, and with $b-2$ of order $N^{-1/3}$ the scaled largest
eigenvalues form a well defined parameter dependent state. Recent works by
Bloemendal and Vir\'ag [BV], and Mo, have quantified this parameter dependent state for 
real Wishart matrices from different viewpoints, and the former authors have done similarly for the
spiked Wishart $\beta$-ensemble. The latter is defined in terms of certain random bidiagonal matrices.
We use a recursive structure to give an alternative construction of the spiked and more generally the general variance Wishart $\beta$-ensemble, and we give the exact form of the joint eigenvalue PDF
for the two matrices in the recurrence. In the case of 
real quaternion Wishart matrices
($\beta = 4$) the latter is recognised as having appeared in earlier studies on symmetrized last passage percolation, 
allowing the exact form of the scaled distribution of the largest eigenvalue to be given. This extends and simplifies earlier work of Wang, and is an alternative derivation to a result in [BV]. We also use the construction of the spiked Wishart $\beta$-ensemble from
[BV] to give a simple derivation of the explicit form of the eigenvalue PDF.
\end{abstract}

\section{Introduction}
\subsection{Background}
Recently, an outstanding problem in random matrix theory has been solved from two different perspectives \cite{BV10,Mo10}. The problem relates to real Wishart matrices, specified as the ensemble of random matrices of the form $X^T X$, where $X$ is an $n \times N$ real Gaussian matrix with distribution proportional to
\begin{equation}\label{1.1}
\exp \Big ( - {N \over 2} {\rm Tr} ( X^T X \Sigma^{-1}) \Big ).
\end{equation}
The case of interest is when the $N \times N$ covariance matrix $\Sigma$ is of the spiked form
\begin{equation}\label{1.2}
\Sigma = {\rm diag} (b, 1^{N-1}).
\end{equation}
Here the notation $1^{N-1}$ denotes the eigenvalue $1$ repeated $N-1$ times, and the eigenvalue
$b$ corresponds to the spike.

With $\gamma = n/N \ge 1$ and $n,N$ large it was shown by Baik and Silverstein \cite{BS06} that for
$b > 1 + \sqrt{\gamma}$ the largest eigenvalue separates from the remainder of the spectrum, which is otherwise  supported on $((1 - \sqrt{\gamma})^2,(1 + \sqrt{\gamma})^2)$. A simplified derivation of this fact can be found in \cite[Prop.~2.4]{BFF09}, and some recent generalisations are given in
\cite{BN09}. This same effect holds for spiked Wishart matrices with complex entries
\cite{BBP05}. There it has been explicitly demonstrated that for  large
$N$, and with $b - (1 + 1/\sqrt{\gamma})$  of order $N^{-1/3}$, the scaled largest eigenvalues form a parameter dependent state at the onset of the eigenvalue separation.
It was conjectured in \cite{BBP05} that for these same scaling parameters, spiked real Wishart matrices similarly exhibit a parameter dependent state. The outstanding problem has been to quantify this state.

Before discussing the two recent works which solve this problem, let us say some more about the complex case. For this the parameter dependent state was shown in \cite{FR02b} (in the case $n = N - 1$), in \cite{BBP05} for general $\gamma > 1$, and in
\cite{On08} for $\gamma < 1$, to be a determinantal point process with correlation kernel independent of $\gamma$. Technically, the correlation kernel was shown to be a rank one perturbation of the familiar Airy kernel \cite{Fo93a}. This explicit form was used in  \cite{Ba05a} to express the distribution of the largest eigenvalue in terms of a member of the Lax pair for the Hasting-McLeod solution of the Painlev\'e II equation, the latter being specified as the transcendent
$q(s)$ satisfying
\begin{equation}\label{qs}
q'' = s q + 2 q^3, \qquad q(s) \mathop{\sim}\limits_{s \to \infty} {\rm Ai}(s).
\end{equation}
 As an application, a known correspondence \cite{Jo99a} between the eigenvalues of complex Wishart matrices and last passage times for a directed percolation model based on the Robinson-Schensted-Knuth correspondence allowed for an interpretation of these results to be given within a statistical mechanics setting.

The most prominent application of Wishart matrices is to principal component analysis in multivariate statistics. There $X$ in (\ref{1.1}) corresponds to a data matrix for $n$ distinct measurements of $N$ different quantities, and thus has real entries. We know from explicit results obtained in the null case for $\gamma > 1$ \cite{Jo01} (the null case refers to $\Sigma = \mathbb I_N$) that the scaled largest eigenvalues now form a Pfaffian rather than determinantal point process. The distribution of the scaled largest eigenvalue again involves the Hasting-McLeod solution of the Painlev\'e II equation (\ref{qs}), but is distinct from that in the complex case 
(see e.g.~\cite[\S 9.7]{Fo10}).

 More recently this so called soft edge state, for both the real ($\beta = 1$) and complex ($\beta = 2$) cases, has further been characterised \cite{ES06,RRV06} in terms of the smallest eigenvalues of the
stochastic Airy operator 
\begin{equation}\label{B1}
- {d^2 \over dx^2} + x + {2 \over \sqrt{\beta}} B'(x), \qquad x \ge 0.
\end{equation}
Here $B(x)$ denotes standard Brownian motion and the eigenfunctions are subject to a Dirichlet boundary condition at $x=0$. This in turn allows for a diffusion characterisation of the distribution of the largest eigenvalue.

In two recent works --- by Bloemendal and Vir\'ag \cite{BV10} and Mo \cite{Mo10} --- the problem of quantifying the soft edge, parameter dependent state for spiked real  Wishart matrices has been solved. The characterisations are very different, in keeping with the two distinct characterisations of the scaled largest eigenvalues revised above in the null case. 

Consider first the work \cite{BV10}. With
$$
- {\sqrt{\gamma} \over (n^{-1/2} + N^{-1/2})^{2/3}} \Big ( b - (1 + 1/\sqrt{\gamma}) \Big ) \to w
$$
as $n,N \to \infty$, and the scaling of the large eigenvalues $\lambda_1 > \lambda_2 > \cdots$ of
$X^T X$ 
$$
 {1 \over \sqrt{n N}} {1 \over (n^{-1/2} + N^{-1/2})^{4/3}}
\Big ( \lambda_k - (\sqrt{n} + \sqrt{N})^2 \Big ) =: Y_k,
$$
it is proved that $\{Y_k\}$ form a well defined parameter dependent state.
The latter is again specified by the smallest eigenvalues of the stochastic Airy operator, but now
with the eigenfunctions satisfying the boundary condition $\psi'(0) = w \psi(0)$. Furthermore, it is shown that the distribution function for the largest eigenvalue, $F_{\beta, w}(x)$ say, is the unique bounded solution to the boundary value problem
\begin{eqnarray}\label{B2}
&&{\partial F \over \partial x} + {2 \over \beta} {\partial^2 F \over \partial w^2} +
(x - w^2) {\partial F \over \partial w}  = 0 \label{B2}  \nonumber \\
&&F(x,w) \to 1 \qquad {\rm as} \: \: x,w \to \infty \: \: {\rm together} \nonumber \\
&&F(x,w) \to 0 \qquad {\rm as} \: \: w \to - \infty \; \; {\rm with} \: \:  x \le x_0 < \infty,
\end{eqnarray}
where $x_0$ is fixed.  An essential step, following \cite{Si85,DE02} is to use Householder transformations to reduce $X^T$ to the $N \times N$ bidiagonal form
\begin{equation}\label{Bd}
B_\beta^T :=
\begin{bmatrix} \sqrt{b} \chi_{\beta n} & & &  \\
\chi_{\beta (N - 1)} & \chi_{\beta (n - 1)} & &  \\
 & \chi_{\beta (N - 2)} & \chi_{\beta (n - 2)} &  & \\
 & \ddots & \ddots &  & \\
& & \chi_\beta & \chi_{\beta (n - N + 1)} 
\end{bmatrix}
\end{equation}
with $\beta = 1$.
Here $\chi_n^2$ refers to the particular gamma distribution $\Gamma[n/2,2]$
(in general $\Gamma [s,\sigma]$ is specified by the PDF proportional to
$x^{s-1} e^{-s/\sigma}$, $x > 0$), it has been assumed for definiteness that $n \ge N$ and some zero columns which do not effect the non-zero eigenvalues of $X^T X$ 
have been removed.

It is the Pfaffian point process characterisation of the null case that is generalised in \cite{Mo10}. Here knowledge of the joint eigenvalue PDF in the finite system is essential. With $N$ even,
this is shown to be proportional to
\begin{equation}\label{Ma}
\prod_{j=1}^N \lambda_j^{(n-N-1)/2}  e^{- \lambda_j/2}
\prod_{1 \le j < k \le N} (\lambda_j - \lambda_k)
\int_{\Gamma} e^t  \prod_{j=1}^N \Big ( t - {b - 1 \over 2b} \lambda_j
\Big )^{-1/2} \, dt,
\end{equation}
for $\Gamma$ a simple closed contour enclosing the branch points of the integrand.
Next, using integration methods based on skew orthogonal polynomials (see \cite[Ch.~6]{Fo10})
a Pfaffian formula is given for the correlations, and the distribution of the largest eigenvalue is expressed in terms of the corresponding Fredholm determinant. 

The starting point of \cite{Mo10} is the expression for the eigenvalue PDF of real Wishart matrices
\begin{equation}\label{Ma1}
{1 \over C} \prod_{j=1}^N \lambda_j^{(n-N-1)/2}  e^{- \lambda_j/2}
\int_{O(N)} e^{- {1 \over 2} {\rm Tr} (O X^T X O^T \Sigma^{-1})} (O^T d O),
\end{equation}
where $C$ is the normalization and $(O^T d O)$ is the Haar measure for real
orthogonal matrices.  In the case of complex Wishart matrices, the corresponding formula involves an average over $U(N)$ rather
than $O(N)$. According to the well known Harish-Chandra/ Itzykson-Zuber formula
(see e.g.~\cite[Prop.~11.6.1]{Fo10}) an evaluation in terms of determinants is possible.
However, until \cite{Mo10}, it was not known that the $O(N)$ matrix integral admitted a
tractable evaluation.

\subsection{An alternative viewpoint and outline}
We have seen that two seemingly distinct viewpoints have led to the quantification of the parameter dependent state formed at the spectrum edge for spiked real Wishart matrices in the critical regime. In this paper we will emphasize a third viewpoint. The idea, initiated in \cite{FR02b}, is to consider the spiking as a perturbation, and to focus attention on the joint eigenvalue distribution of the perturbed and unperturbed matrices. This follows naturally from the recurrence
\begin{equation}\label{Y0}
X X^T = \tilde{X} \tilde{X}^T + b \vec{x} \vec{x}^T
\end{equation}
for $n \times N$ matrices $X$ specified by (\ref{1.1}) and (\ref{1.2}), where $\tilde{X}$ is an
$n \times (N-1)$ matrix of standard Gaussians obtained from $X$ by deleting its first column, and $\vec{x}$ is an $n$-component vector of standard Gaussians.

We will use this formalism to give an alternative construction of the spiked 
 Wishart $\beta$-ensemble, specified as the random matrices $B_\beta^T B_\beta$,
 with $B_\beta$ the $N \times N$ bidiagonal matrix (\ref{Bd}).  This in turn relies on knowledge of the eigenvalue PDF for the Wishart $\beta$-ensemble as specified in terms of
 (\ref{Bd}). We begin in Section 2 by showing how to deduce the eigenvalue PDF (\ref{Ma}) from the bidiagonal matrix (\ref{Bd}) with $\beta = 1$. Our derivation applies for all $\beta > 0$, so
 we are able to give the $\beta$ generalization of (\ref{Ma}). This is given in (\ref{PB}) below.
 
 In Section 3 we use known results from \cite{FR02b} to compute the joint eigenvalue PDF of the
non-zero eigenvalues of the random matrix pair $( \tilde{X} \tilde{X}^T, X X^T)$, as related by (\ref{Y0}) but with $\tilde{X} \tilde{X}^T$ replaced by diag$ \, \tilde{X} \tilde{X}^T$, in the case that the non-zero eigenvalues of $\tilde{X} \tilde{X}^T$ have PDF
proportional to 
\begin{equation}\label{Yy}
\prod_{l=1}^{N-1} y_l^{\beta(n - N + 2)/2 - 1} e^{-y_l/2}
\prod_{1 \le j < k \le N-1} (y_j - y_k)^{\beta}.
\end{equation}
Results from \cite{FR02b} tell us that this joint eigenvalue PDF can be realized as the zeros
of two polynomials generated recursively from a three term recurrence. We give the explicit
form of the matrix eigenvalue problem implied by the recurrences.
Although different to the tridiagonal matrix eigenvalue problem for $B_\beta^T B_\beta$,
it similarly involves only $2N-1$ independent entries.
We also take up
 the problem of integrating over the eigenvalues of $\tilde{X} \tilde{X}^T $,
with the aim of showing that the non-zero eigenvalues of $X X^T$ have the same PDF 
as found in Section 2 for the spiked 
 Wishart $\beta$-ensemble, thus providing an alternative construction of this ensemble.
 We remark that this construction also allows for a $\beta$-generalisation of the 
 general variance Wishart ensemble, and the corresponding eigenvalue PDF. From the latter,
 for the special value $\beta(n-N+1)/2-1 = 0$, it is possible to show that the probability of no
 eigenvalues in $(0,s)$ has a simple exponential distribution.
 
 With $n < N$, $\tilde{X} \tilde{X}^T$ no longer has any zero eigenvalues. This setting is studied
 in Section 4. 
In the case $\beta = 4$, and for a special $n$ we obtain a joint eigenvalue PDF 
 proportional to
  \begin{equation}\label{xy1}
  e^{-\sum_{j=1}^N ( y_j + (\lambda_j - y_j)/b)/2}
  \prod_{1 \le i < j \le N} (\lambda_i - \lambda_j) (y_i - y_j)
  \prod_{i,j=1}^N |\lambda_i - \lambda_j|,
  \end{equation}
 subject to the interlacing
 \begin{equation}\label{xy}
 \lambda_1 > y_1 > \lambda_2 > y_2 > \cdots > \lambda_N > y_N \ge 0.
 \end{equation}
 The corresponding parameter dependent soft edge correlations were calculated as
 a Pfaffian in \cite{FR02}. By the universality results of \cite{BV10} these same correlations
 must hold for all cases of the $\beta = 4$ spiked Wishart matrices (i.e.~for all choices of
 $n$ and $N$ in (\ref{Bd}) provided they both go to infinity). Moreover, results from
 \cite{BR01a} tell us that the scaled distribution of the largest eigenvalue with PDF
 (\ref{xy}) can be written in terms of the same member of the Lax pair for the Hastings-McLeod
 solution of the Painlev\'e II equation as known for the complex case \cite{Ba05a}. And universality tells us that this result must persist for all cases of the  $\beta = 4$ spiked Wishart matrices.
 An alternative derivation of this fact was given by Bloemendal and Vir\'ag \cite{BV10}, who showed that the distribution satisfies (\ref{B2}).
 Our results of this section  extend the results of Wang \cite{Wa09}, who considered a particular
 value of the parameter only.

\section{Eigenvalue PDF for the spiked Wishart $\beta$-ensemble}
\setcounter{equation}{0}
By the spiked Wishart $\beta$-ensemble we refer to the tridiagonal matrices
$B_\beta^T B_\beta$, with $B_\beta$ the $N \times N$ bidiagonal matrix (\ref{Bd}).
For $\beta = 1,2 $ and 4 we know that this tridiagonal matrix corresponds to a unitary similarity
transformation of the spiked real, complex and real quaternion Wishart matrices, and so
shares the same eigenvalue PDF. Here we seek the eigenvalue PDF of $B_\beta^T B_\beta$
for general $\beta > 0$. The resulting expression, (\ref{PB}), reproduces in the case $\beta = 1$, $N$ even (\ref{Ma})
as first derived in \cite{Mo10}; for $\beta = 1$, 2 and 4 it agrees with results obtained in \cite[Appendix A]{Wa10}
and it generalizes those results in the case of $\beta = 1$, $N$ odd.
In the case $b=1$, the task has been carried out in \cite{DE02}. We can adapt the workings of that calculation to the general $b > 0$ case.

\begin{proposition}
Define the branch of $z^{-\beta/2}$ by the equation
$$
z^{-\beta/2} = {1 \over \Gamma(\beta/2)} \int_0^\infty t^{\beta/2 - 1} e^{-t z} \, dt,
$$
or equivalently as $z^{-\beta/2} = e^{-(\beta/2) \log z}$, where $\log z$, $z \notin \mathbb R^{-}$ is defined by its principal branch.
The tridiagonal matrix $B_\beta^T B_\beta$ has eigenvalue PDF proportional to 
\begin{equation}\label{PB}
\prod_{j=1}^N \lambda_j^{\beta (n - N + 1)/2 - 1} e^{- \lambda_j/2}
\prod_{1 \le j < k \le N} (\lambda_j - \lambda_k)^\beta
\int_{-\infty}^\infty e^{i t} \prod_{j=1}^N \Big ( i t - {b -1 \over 2 b} \lambda_j \Big )^{-\beta/2}
\, dt.
\end{equation}
\end{proposition}

\noindent
Proof. \quad Let us write
\begin{equation}\label{Ba}
 B_\beta^T :=
\left [ \begin{array}{cccc}
x_n & & & \\
y_{N-1} & x_{n-1} & & \\
&\ddots & \ddots &  \\
&& y_1 & x_{n-N+1}  \end{array} \right ].
\end{equation}
Then, according to the definition (\ref{Bd}), the probability measure
$P(B_\beta) (d B_\beta)$ has, up to proportionality, the factorization
\begin{equation}\label{Bb}
\Big ( P(B_\beta) (d B_\beta) \Big ) \Big |_{b = 1} e^{(1 - 1/b) x_n^2/2}.
\end{equation}

Let us denote by $\{\lambda_j\}_{j=1,\dots,N}$ the (ordered) eigenvalues of $B_\beta$,
and by $\{q_j\}$ the first component of the corresponding (normalized) eigenvector.
The working of \cite{DE02} (see also \cite[proof of Prop. 3.10.1]{Fo10}) tells us that
in terms of these variables $P(B_\beta) (d B_\beta)  |_{b = 1}$ is proportional to 
\begin{equation}\label{Bc}
\prod_{j=1}^N \lambda_j^{\beta(n-N+1)/2 - 1} e^{- \lambda_j/2}
\prod_{1 \le j < k \le N} (\lambda_j - \lambda_k)^\beta
\prod_{i=1}^N q_i^{\beta - 1} \delta \Big ( \sum_{j=1}^N q_j^2 - 1 \Big ) \,
(d \vec{\lambda}) (d \vec{q}),
\end{equation}
where $\delta( \cdot)$ denotes the Dirac delta function. Furthermore, if we write
$$
B_\beta^T B_\beta =
\left [ \begin{array}{ccccc}
a_N &b_{N-1} & & & \\
b_{N-1} & a_{N-1} &b_{N-2} & &\\
&\ddots & \ddots &\ddots & \\
&& b_2 & a_{2} & b_1 \\
&& & b_1 & a_1 \end{array} \right ],
$$
then we see from (\ref{Ba}) that $a_N = x_n^2$. But we also know \cite[proof of Prop. 1.9.3]{Fo10} that   $a_N = \sum_{j=1}^N q_j^2 \lambda_j$. Hence, substituting $x_n^2 =
 \sum_{j=1}^N q_j^2 \lambda_j$ in (\ref{Bb}) we see from (\ref{Bc}) that our remaining task is to
 compute
 \begin{equation}\label{Bdi}
 \int_{(\mathbb R^+)^N} \prod_{i=1}^N q_i^{\beta - 1} \delta  \Big ( \sum_{j=1}^N q_j^2 - 1 \Big )
  e^{(1 - 1/b) \sum_{j=1}^N q_j^2 \lambda_j /2} \, (d \vec{q}).
  \end{equation}
  
  Introducing the integral form of the delta function
  $$
  \delta \Big ( \sum_{j=1}^N q_j^2 - 1 \Big ) = \lim_{\epsilon \to 0^+}
  {1 \over 2 \pi} \int_{0}^\infty
  e^{i t ( 1 -  \sum_{j=1}^Nq_j^2  )} e^{ - \epsilon t^2} \, dt
  $$
  and supposing temporarily that $b < 1$ so the coefficient on the exponential in (\ref{Bd})
  is negative, we see that (\ref{Bdi}) is equal to
  $$
  {1 \over 2 \pi} \int_{-\infty}^\infty e^{i t} \prod_{j=1}^N
  \Big ( \int_{-\infty}^\infty  q^{\beta - 1} e^{- q^2 ( it  + (1/2)(1/b - 1) \lambda_j)} \, dq  \Big ) dt.
  $$
  Evaluating the integral, up to proportionality this reduces to 
  \begin{equation}\label{Be}
  \int_{-\infty}^\infty e^{i t} \prod_{j=1}^N \Big ( i t - {b -1 \over 2 b} \lambda_j \Big )^{-\beta/2}
\, dt,
\end{equation}
and we see furthermore that the restriction to $b< 1$ can now be relaxed.
Multiplying the eigenvalue dependent factors of (\ref{Bc}) with (\ref{Be}) gives
(\ref{PB}). \hfill $\square$

\section{An alternative construction of the spiked Wishart $\beta$-ensemble}
\setcounter{equation}{0}
\subsection{Joint eigenvalue PDF for $( \tilde{X} \tilde{X}^T, X X^T)$}
We begin by giving the derivation of the recurrence (\ref{Y0}). With $X$ distributed as in
(\ref{1.1}), set $X = Y \Sigma^{1/2}$. We see that $Y^TY$ is then distributed as a real Wishart
matrix with variance matrix equal to the identity ($\Sigma = \mathbb I_N$). With $\Sigma$ as
in (\ref{1.2}) it then follows that each element in the first column of $X$ has distribution
N$[0,\sqrt{b}]$ (i.e.~is a zero mean, standard deviation  $\sqrt{b}$ Gaussian), and all other elements are
distributed independently as N$[0,1]$. Hence the matrix product 
$X X^T$ can be factorized according to the RHS of  (\ref{Y0}).
An analogous factorization holds for $X$ having complex elements ($\beta = 2$) or
real quaternion elements ($\beta = 4$). Moreover in each case, by the invariance of the
distribution of a Gaussian vector under conjugation by a unitary matrix, we have that
\begin{equation}\label{ePDF}
{\rm ePDF} \, X X^\dagger = {\rm ePDF} \, \Big ({\rm diag} \, (\tilde{X} \tilde{X}^\dagger) + b \vec{x} \vec{x}^T \Big )
\end{equation}
where with $q_i := |x_i|^2$ (the squared modulus of the entries of $\vec{x}$), we have that each $\{q_i\}$
is distributed according to the gamma distribution $\Gamma[\beta/2,2]$ ($\beta = 1,2$ and 4).
In the case $\beta = 4$ of real quaternion entries, one has that all eigenvalues are doubly degenerate.
The notation ePDF in (\ref{ePDF}) refers
to the eigenvalue PDF.

The equation (\ref{ePDF}), valid for $\beta = 1,2$ and 4, suggests the recursive construction of $n \times n$ matrices
$Y_{n,N}$, depending also on a parameter $N$ ($n \ge N$), according to
\begin{equation}\label{3.1}
Y_{n,N} = {\rm diag} \, Y_{n,N-1} + b \vec{x} \vec{x}^T,
\end{equation}
where the random vector $\vec{x}$ has components $x_i$ such that $q_i := |x_i|^2$ is distributed according to the gamma distribution
$\Gamma[\beta/2,2]$, and with initial condition $Y_{n,0} = 0_{n \times n}$. Note that $Y_{n,m}$ has $m$ nonzero eigenvalues.
Furthermore, it is a standard result \cite[eq.~(3.16)]{Fo10} that in the case $b=1$ and $\beta = 1$, 2 and 4 
the joint PDF of these eigenvalues is given by
(\ref{Yy}) with $N-1 = m$. 
We take up the the problem of computing the
joint distribution of the eigenvalues of $Y_{n,N-1}$ and $Y_{n,N}$, 
under the assumption that (\ref{Yy}) is the eigenvalue PDF of $Y_{n,N-1}$ for general $\beta > 0$.
with the two matrices related by (\ref{ePDF}). 

Let the non-zero eigenvalues of $Y_{n,N-1}$ be denoted by $\{y_i\}_{i=1,\dots,N-1}$. It is a simple exercise to show that the secular equation for the eigenvalue problem implied by
(\ref{ePDF}) is
\begin{equation}\label{Y4}
0 = 1 + b \Big ( - {q_0 \over \lambda} + \sum_{j=1}^{N-1} {q_j \over y_j - \lambda} \Big ),
\end{equation}
where $q_j \mathop{=}\limits^{\rm d} \Gamma[\beta /2,2]$ $(j=1,\dots,N-1)$ and
$q_0 \mathop{=}\limits^{\rm d} \Gamma[\beta (n-N+1)/2,1/2]$.
Furthermore, we know from \cite[Cor.~3]{FR02b} that the PDF of the roots of this equation, and thus the conditional PDF of the non-zero eigenvalues $\{\lambda_i\}_{i=1,\dots,N}$ of $\tilde{X} \tilde{X}^T$,
is proportional to 
\begin{eqnarray}\label{Y5}
 &&
 \prod_{j=1}^N \lambda_j^{\beta (n - N + 1)/2 - 1} e^{- \lambda_j/ 2b}
\prod_{l=1}^{N-1}  y_l^{-\beta(n - N + 2)/2 + 1}  e^{- y_l/2b} 
\nonumber \\
 && \quad \times
{ \prod_{1 \le j < k \le N} (\lambda_j - \lambda_k) \over  \prod_{1 \le j < k \le N - 1} (y_j - y_k)^{\beta-1} }
 \prod_{i=1}^N \prod_{j=1}^{N-1} |\lambda_i - y_j|^{\beta/2 - 1},
 \end{eqnarray}
 subject to the interlacing (\ref{xy}) with $y_N:=0$. Our sort result can now be deduced.
 
 \begin{proposition}
 Let $Y_{n,N-1}$ and $Y_{n,N}$ be related by (\ref{ePDF}), and suppose that the non-zero
 eigenvalues of  $Y_{n,N-1}$ are denoted  $\{y_i\}_{i=1,\dots,N-1}$ and have PDF
 given by (\ref{Yy}). With $\{\lambda_i\}_{i=1,\dots,N}$ denoting the non-zero eigenvalues of
 $Y_{n,N}$, we have that the joint eigenvalue PDF of both sets of non-zero eigenvalues is proportional to 
 \begin{eqnarray}\label{30.4}
&& \prod_{j=1}^N \lambda_j^{\beta ( n - N + 1)/2 - 1} e^{-\lambda_j/2b}
\prod_{l=1}^{N-1}   e^{-(1-1/b)y_l/2} \nonumber \\
&& \quad \times \prod_{1 \le j < k \le N-1} (y_j - y_k)
 \prod_{1 \le j < k \le N} (\lambda_j - \lambda_k)
 \prod_{i=1}^N \prod_{j=1}^{N-1} |\lambda_i - y_j|^{\beta/2 - 1},
 \end{eqnarray}
 subject to the interlacing (\ref{xy}) with $y_N:=0$.
 \end{proposition}
 
 \noindent
 Proof. \quad The joint PDF is given by the product of the conditional PDF for
 $\{\lambda_i\}_{i=1,\dots,N}$ given $\{y_i\}_{i=1,\dots,N-1}$, times the PDF of $\{y_i\}_{i=1,\dots,N-1}$.
 Thus we need to multiply together (\ref{Y5}) and (\ref{Yy}), and (\ref{30.4}) results.
 \hfill $\square$

\smallskip
What is the marginal distribution of $\{\lambda_i\}_{i=1,\dots,N}$? In the cases $\beta =1$, 2 and 4,
the construction (\ref{ePDF}) is equivalent the distribution of $X^\dagger X$ being given by the
spiked Wishart distribution (\ref{1.1}) and (\ref{1.2}). Furthermore the non-zero eigenvalues
of $X X^\dagger$ are the same as the non-zero eigenvalues of $X^\dagger X$. Hence it must be in these cases at least,
$\{\lambda_i\}_{i=1,\dots,N}$ has PDF (\ref{PB}). We would like to show that this remains true for
general $\beta > 0$. Our task then is to integrate over $\{y_i\}_{i=1,\dots,N-1}$ in
(\ref{30.4}), and show that (\ref{PB}) results. This can be accomplished by the use of Jack polynomial theory \cite[Ch.~12\&13]{Fo10}. But before taking on this task, we will make note
of a realization of (\ref{30.4}) in terms of a generalised eigenvalue problem.

\subsection{Relationship to a bidiagonal generalised eigenvalue problem}
Let
\begin{eqnarray}\label{abc}
&& a_j \mathop{=}\limits^{\rm d} \Gamma[(N-j)\beta/2 + \alpha_0 + 1, 2] \: \:
(j=1,\dots,N-1) \quad a_N \mathop{=}\limits^{\rm d} \sqrt{b} \Gamma[\alpha_0+1,2] \nonumber \\
&& b_j  \mathop{=}\limits^{\rm d} \Gamma[j \beta/2,2] \: \:
(j=1,\dots,N-2) \qquad \quad b_{N-1}  \mathop{=}\limits^{\rm d} \sqrt{b}  \Gamma[(N-1)\beta/2,2]
\end{eqnarray}
where $\alpha_0 := \beta (n-N+1)/2 - 1$, and set too $b_0 := 0$.
Then results from \cite[Section 5.2]{FR02b} tell us that with the monic random polynomials
$\{B_j(x)\}_{j=0,\dots,N}$ defined by the three term recurrence
\begin{equation}\label{B}
B_j(x) = (x - a_j) B_{j-1}(x) - b_{j-1} x B_{j-2}(x) \: \: (j=1,\dots,N),
\end{equation}
we have that the joint PDF of the zeros of $(B_N(x), B_{N-1}(x))$ is given by (\ref{30.4}).

In general (see e.g.~\cite{SV02}) the recurrence (\ref{B}) is satisfied by the characteristic polynomials
$B_j(x) = \det (x M_j - L_j)$ where $L_j$ and $M_j$ are the top $j \times j$ blocks of the
bidiagonal matrices
$$
L:= \begin{bmatrix}a_1 & 1 & & &\\
 & a_2 & 1 & &\\
& & \ddots & \ddots  &\\
& & & a_{N-1} & 1 \\
& & &  & a_N \end{bmatrix}, \quad
 M:= \begin{bmatrix}1 &  & & &\\
-b_1 & 1 &  & &\\
 & -b_2 & 1 && \\
 &  & \ddots & \ddots  &\\
& & &  -b_{N-1} & 1 \end{bmatrix}
 $$
 We see from the specification of the entries of (\ref{abc}) that the spike $b$ effects only the
 single entries $a_N$ and $b_{N-1}$ in $L$ and $M$ respectively, which is analogous to
 how $b$ enters (\ref{Bd}).
 
 An open problem is to obtain the stochastic characterisation of the soft edge spiked
 Wishart $\beta$-ensemble starting from the generalised eigenvalue problem 
 $L \vec{v} = \lambda M \vec{v}$.

\subsection{Jack polynomials and hypergeometric functions}
The conditional PDF (\ref{Y5}) is a special case of the Dixon-Anderson density
\cite[eq.~(4.11)]{Fo10}. Another special case is the conditional PDF for
$\{y_i\}_{i=1,\dots,N-1}$  given $\{\lambda_i\}_{i=1,\dots,N}$ 
\begin{equation}\label{DA}
{\Gamma(N\beta/2) \over (\Gamma(\beta/2))^N}
{\prod_{1 \le j < k \le N - 1} (y_j - y_k) \over
\prod_{1 \le j < k \le N} (\lambda_j - \lambda_k)^{\beta - 1}}
\prod_{i=1}^{N-1} \prod_{j=1}^N | y_i - \lambda_j|^{\beta/2 - 1}
\end{equation}
subject to the interlacing (\ref{xy}) with $y_N :=0$. Let this be referred to as DA${}_N(\beta/2)$.

Intimately related to (\ref{DA}) are the symmetric Jack polynomials $P_\kappa(z;\alpha)$,
where $z = (z_1,\dots,z_N)$, $\kappa$ denotes a partition of length less than or equal to
$N$ (we write $\ell(\kappa) \le N$), and $\alpha$ is a parameter. The Jack polynomials can
be specified as the polynomial eigenfunctions of the differential operator
$$
\sum_{j=1}^N \Big ( z_j {\partial \over \partial z_j} \Big )^2 +
{2 \over \alpha} \sum_{1 \le j < k \le N}
{z_j + z_k \over z_j - z_k} \Big ( {\partial \over \partial z_j} - {\partial \over \partial z_k} \Big ),
$$
with leading term given by the monomial symmetric function $m_\kappa(z)$ (see
\cite[\S 12.6]{Fo10} for more details). Thus with the generalised Pochhammer symbol
specified by
\begin{equation}\label{uk}
[u]_\kappa^{(\alpha)} = \prod_{j=1}^N {\Gamma(u - (j-1)/\alpha + \kappa_j)
\over \Gamma(u - (j-1)/\alpha)},
\end{equation}
we have \cite[eq.~(12.209)]{Fo10}
\begin{equation}\label{DAP}
P_\kappa(\lambda;2/\beta) =
{[\beta N /2]_\kappa^{(2/\beta)} \over [\beta (N - 1)/2]_\kappa^{(2/\beta)}}
\langle P_\kappa(y;2/\beta) \rangle_{{\rm DA}_N(\beta/2)},
\end{equation}
valid for $\ell(\kappa) \le N - 1$.

Let us define the quantity $d_\kappa'$ as in \cite[eq.~(12.60)]{Fo10} (it's precise value plays no explicit role in the following), and use this in the definition of the renormalized Jack polynomials
$$
C_\kappa(z;\alpha) = {\alpha^{|\kappa|} |\kappa|! \over d_\kappa'}
P_\kappa(z;\alpha).
$$
The generalized hypergeometric functions based on Jack polynomials are then specified by
\begin{eqnarray}\label{pqF}
&& {}_p^{} F_q^{(\alpha)}(a_1,\dots,a_p;b_1,\dots,b_q; z) 
 = \sum_{\kappa} {1 \over |\kappa|!}
{[a_1]_\kappa^{(\alpha)} \cdots [a_p]_\kappa^{(\alpha)} \over
[b_1]_\kappa^{(\alpha)} \cdots [b_q]_\kappa^{(\alpha)} }C_\kappa(z;\alpha).
\end{eqnarray}
Important for our present purposes is the fact that  \cite[eq.~(13.3)]{Fo10}
\begin{equation}\label{00F}
{}_0^{} F_0^{(\alpha)}(z) = e^{\sum_{j=1}^N z_j}.
\end{equation}

The use and relevance of the generalized hypergeometric functions reveals itself upon
multiplying both sides of (\ref{DAP}) by
$$
{ [\beta (N - 1)/2]_\kappa^{(2/\beta)} \over [ \beta N /2]_\kappa^{(2/\beta)} }
{(2/\beta)^{|\kappa|} \over d_\kappa'} \Big ( {1 \over 2} \Big ({1 \over b} - 1 \Big ) \Big )^{|\kappa|}
$$
and making use of (\ref{pqF}) with $p=q=1$ on the LHS. On the RHS we first use the fact
that the Jack polynomials are homogeneous of degree $|\kappa|$, and so for $c$ a scalar
$P_\kappa(zc;\alpha) = c^{|\kappa|} P_\kappa(z;\alpha)$, then use (\ref{00F}). We thus obtain
the following corollary of (\ref{DAP}).

\begin{proposition}
For $\ell(\kappa) \le N - 1$ and $x := (x_1,\dots,x_N)$ we have
\begin{eqnarray}\label{DAc}
\Big \langle  e^{(1/b - 1) \sum_{j=1}^{N-1} y_j/2} \Big \rangle_{{\rm DA}_N(\beta/2)}
&= & {}_1 F_1^{(2/\beta)} ( \beta (N-1)/2;\beta N/2;
(1/b - 1) z/2) \nonumber \\
&= & e^{(1/b - 1) \sum_{j=1}^N z_j/2}
{}_1^{} F_1^{(2/\beta)}(\beta/2;\beta N/2; (1 - 1/b) z/2).
\end{eqnarray}
\end{proposition}

\noindent
Proof. \quad It remains to explain the second line. This follows from a generalisation of the second
Kummer identity \cite[(13.16)]{Fo10}, which states that
$$
{}_1^{} F^{(\alpha)}_1(a;c;z) =
e^{\sum_{j=1}^N z_j}
{}_1^{} F_1^{(\alpha)}(c-a;c;-z).
$$
\hfill $\square$

Comparing the explicit form of DA${}_N(\beta/2)$ (\ref{DA}) with the joint PDF (\ref{30.4}), it
follows that the marginal distribution of $\{\lambda_i\}_{i=1,\dots,N}$ is proportional to
\begin{eqnarray}\label{Fbb}
&& \prod_{j=1}^N \lambda_j^{\beta (n - N + 1)/2 - 1} e^{- \sum_{j=1}^N \lambda_j/2}
\prod_{1 \le j < k \le N} (\lambda_j- \lambda_k)^\beta
{}_1 F_1^{(2/\beta)} (\beta/2;\beta N/2;(1 - 1/b) \lambda/2 ).
\end{eqnarray}
Comparison of (\ref{Fbb})
with (\ref{PB}) shows that our remaining task is to show that for $c$ a scalar
\begin{equation}\label{Fbc}
{}_1^{} F_1^{(2/\beta)} (\beta/2;\beta N/2; c\lambda)
\propto \int_{-\infty}^\infty e^{i t} \prod_{j=1}^N \Big ( i t -  c \lambda_j \Big )^{-\beta/2}
\, dt.
\end{equation}

For this purpose, we begin by observing from (\ref{pqF}) and (\ref{uk}) that in general
${}_1 F_1^{(2/\beta)}(\beta/2;b;z)$ is very special. Thus the only partitions giving a non-zero
contribution to the sum (\ref{pqF}) are of the form $(k,0^{N-1})$, and so the summation
is one-dimensional. In the case $b = \beta N/2$, as is the case in (\ref{Fbc})
 there is a further special feature, relating to the particular generalized hypergeometric
 function based on two sets of variables \cite[eq.~(13.20)]{Fo10}
 \begin{equation}\label{Fg}
 {}_0^{} {\mathcal F}_0^{(2/\beta)}(x;y) := \sum_\kappa {C_\kappa^{(\alpha)}(x) C_\kappa^{(\alpha)}(y) \over |\kappa|!
 C_\kappa^{(\alpha)}(1^N)},
 \end{equation}
 where $x := (x_1,\dots,x_n)$ and $y := (y_1,\dots,y_n)$.
 To see the relation, note that for $\kappa = (k,0^{N-1})$ we have 
 $$
 C_\kappa^{(\alpha)}((c,0^{N-1})) = c^k, \qquad
 C_\kappa^{(\alpha)}(1^N) = {[N/\alpha]_\kappa^{(\alpha)} \over
 [1/\alpha]_\kappa^{(\alpha)} }
 $$
 (for the second formula see e.g.~\cite[eq.~(243)]{Wa10}), while for $\kappa$ with two or more non-zero parts, $C_\kappa^{(\alpha)}((c,0^{N-1})) = 0$. Thus the summation over $\kappa$ in (\ref{Fg}) is
 also one-dimensional, and moreover we have that
 \begin{equation}\label{Fh}
 {}_1^{} F_1^{(2/\beta)} (\beta/2;\beta N/2; cx)
=  {}_0^{} {\mathcal F}_0^{(2/\beta)}(x;(c,0^{N-1})).
\end{equation}

We remark that an alternative derivation of the marginal distribution being given by
(\ref{Fbb}) with the substitution (\ref{Fh}) can be given by using the recursive integration formula
\cite{GK02} (see also \cite[Appendix C]{FR02b})
\begin{eqnarray}\label{3.16b}
&&{}_0^{} {\mathcal F}_0^{(2/\beta)}(\{\lambda\}_{i=1,\dots,N};\{z\}_{i=1,\dots,N}) 
\nonumber \\
&& \qquad =
e^{z_N \sum_{j=1}^N \lambda_j}
\langle e^{- z_N \sum_{j=1}^{N-1} y_j}
 {}_0^{} {\mathcal F}_0^{(2/\beta)}(\{y_i\}_{i=1,\dots,N-1};\{z_i\}_{i=1,\dots,N-1})
 \rangle_{{\rm DA}_N(2/\beta)}.
 \end{eqnarray}
 Also, as noted in \cite{Wa10}, there is a further alternative derivation in the cases $\beta = 1$,
 2 and 4. Thus with $(U^\dagger d U)$ denoting the normalized Haar volume form for
 unitary matrices with real $(\beta = 1)$, complex $(\beta = 2)$ and real
 quaternion $(\beta = 4)$ entries, and $H$, $H^{(0)}$ Hermitian matrices with elements
 from the same field as $U$, eigenvalues $\{\lambda_j\}$, $\{\lambda_j^{(0)}\}$, we know that
 (see e.g.~\cite[eq.~(13.146)]{Fo10})
 \begin{equation}\label{3.17a}
 \int e^{{\rm Tr} (H^{(0)} U^\dagger H U)} (U^\dagger d U) =
 {}_0^{} {\mathcal F}_0^{(2/\beta)}(\lambda^{(0)};\lambda).
 \end{equation}
 This combined with (\ref{Ma1}) and its $\beta = 2$ and 4 analogues gives the result.
 
 Now it has been shown by Wang in \cite[Appendix A]{Wa10} that for $\beta$ even
 \begin{equation}\label{Wang}
 {}_0^{} {\mathcal F}_0^{(2/\beta)}(\lambda;(c,0^{N-1}))
 \propto 
 \int_{\mathcal C} e^{cw} \prod_{j=1}^N {1 \over (w - \lambda_j)^{\beta/2} }
  \, dw
 \end{equation}
 where $\mathcal C$ is a simple closed contour encircling $\{x_j\}$. Supposing temporarily that
 $c < 0$ and $\{x_j\}_{j=1,\dots,N}$ being in the right half plane allows $\mathcal C$ to be
 taken to run along the imaginary axis from $-i \infty$ to $i \infty$, then be closed as an infinite
 half circle in the right half plane. But under the assumption that $c < 0$ there is no contribution to
 the integral along this portion of the contour, due to the integrand vanishing exponentially fast.
 Hence
 \begin{equation}\label{e0F}
 {}_0^{} {\mathcal F}_0^{(2/\beta)}(\lambda;(c,0^{N-1}))
 \propto 
 \int_{-i \infty}^{i \infty}  e^{cw} \prod_{j=1}^N {1 \over (w - \lambda_j)^{\beta/2} }
 \, dw,
 \end{equation}
 and furthermore we can drop the restrictions on $c$ and $\beta$ by analytic continuation
 (the latter requires analytic continuation off the integers; for this we use 
 Carlson's theorem --- see e.g.~\cite[Prop.~4.1.4]{Fo10}).
 
 A useful check on (\ref{e0F}) is to consider that case $N=1$. It follows immediately from 
 the definition (\ref{Fg}) that in this case ${}_0^{} {\mathcal F}_0^{(2/\beta)}(\lambda;c)
 = e^{\lambda c}$. To reclaim this from (\ref{e0F}) we suppose temporarily that
 $0 < \beta/2 < 1$. Then we an change variables $w \mapsto w + x$ to deduce that
 ${}_0^{} {\mathcal F}_0^{(2/\beta)}(\lambda;c) \propto e^{xc}$ as required (the restriction
  $0 < \beta/2 < 1$ can be removed by analytic continuation). We remark that a more
  complicated formula than (\ref{e0F}) has been given in \cite[Appendix A]{Wa10}
  for the continuation of (\ref{Wang}) for general $\beta > 0$. A crucial difference is
  that the latter formula involves the contour $\mathcal C$, whereas 
  the contour in (\ref{e0F}) is along the imaginary axis.
  
  Substituting (\ref{e0F}) in (\ref{Fh}), we see that (\ref{Fbc}) holds as required, and thus
  $\{\lambda_j\}_{j=1,\dots,N}$ as implied by the roots of the equation (\ref{Y4}) indeed realise the
  eigenvalue PDF for the spiked Wishart $\beta$-ensemble.
  
  \subsection{General variance Wishart $\beta$-ensemble}
  Our use of (\ref{ePDF}) has been to perturb the eigenvalue PDF given by (\ref{Yy}),
  and furthermore (\ref{ePDF}) has motivated the recursive construction (\ref{3.1}).
  An extension of the latter is to make the parameter $b$ depend on $N$,
  \begin{equation}\label{3.1a}
Y_{n,N} = {\rm diag} \, Y_{n,N-1} + b_N \vec{x} \vec{x}^T,
\end{equation}
where again $Y_{n,0} = 0_{n \times n}$.
  From the discussion of the first paragraph of \S 3.1, for $\beta = 1,2$ and 4 we must have that
  ePDF$Y_{n,N}$ is proportional to the known eigenvalue PDF for general variance
  Wishart matrices with real ($\beta = 1$), complex ($\beta = 2$) and real quaternion
  ($\beta = 4$) elements. According to (\ref{Ma1}) (appropriately generalized for $\beta = 2$
  and 4), and (\ref{3.17a}), the latter is proportional to
   \begin{equation}\label{cb1}
   \prod_{j=1}^N \lambda_j^{\beta (n-N + 1)/2) - 1} e^{-\lambda_j/2}
   \prod_{1 \le j < k \le N} (\lambda_j - \lambda_k)^\beta
   {}_0^{} {\mathcal F}_0^{(2/\beta)}((b-1)/2b;\lambda),
   \end{equation}
   where $(b-1)/2b := ((b_1-1)/2b_1,\dots,(b_N-1)/2b_N)$. This functional form was proposed
  recently by Wang \cite{Wa10} as a natural $\beta$-generalisation of the
  eigenvalue PDF for the general variance Wishart matrices. We can use (\ref{3.1a}) to give
  a random matrix realization.

  Thus for general $\beta > 0$ the conditional PDF (\ref{Y5}) gives a recurrence for
  ePDF$ Y_{n,N}$. The recursive integration formula (\ref{3.16b}) tells us that
  (\ref{cb1}) is the solution of this recurrence, and thus we can realize (\ref{cb1}) as the
  eigenvalue PDF for this recursively constructed random matrix ensemble.

  Although we have emphasized soft edge scaling in the Introduction, it is worth remarking that
  in the case $\beta(n-N+1)/2 - 1 = 0$ of (\ref{cb1}) (i.e.~when the factors of powers of the
  $\lambda_j$ are not present), there is a very simple formula for the probability of no
  eigenvalues in $(0,s)$. The latter corresponds to the hard edge gap probability,
  $E_\beta(0;(0,s))$ say. According to (\ref{cb1}), in this setting and for an appropriate normalization $C$,
  \begin{eqnarray*}
  &&E_\beta(0;(0,s))  \nonumber \\
  && \quad =  {1 \over C} \int_s^\infty d \lambda_1 \cdots
  \int_s^\infty d \lambda_N \, e^{-\sum_{j=1}^N \lambda_j/2}
  \prod_{1 \le j < k \le N} (\lambda_j - \lambda_k)^\beta
   {}_0^{} {\mathcal F}_0^{(2/\beta)}((b-1)/2b;\lambda) \\ 
  & & \quad = {e^{-Ns/2} \over C} \int_0^\infty d \lambda_1 \cdots
  \int_0^\infty d \lambda_N \, e^{-\sum_{j=1}^N \lambda_j/2}
  \prod_{1 \le j < k \le N} (\lambda_j - \lambda_k)^\beta
   {}_0^{} {\mathcal F}_0^{(2/\beta)}((b-1)/2b;\lambda+s), 
   \end{eqnarray*}
   where $\lambda + s := (\lambda_1 + s,\dots,\lambda_N + s)$.
   But we know that \cite{BF98b}
   $$
    {}_0^{} {\mathcal F}_0^{(2/\beta)}((b-1)/2b;\lambda+s)
    = e^{s \sum_{j=1}^N (b_j-1)/2b_j} {}_0^{} {\mathcal F}_0^{(2/\beta)}((b-1)/2b;\lambda)
    $$
    and hence
    \begin{equation}\label{3ee}
    E_\beta(0;(0,s)) = e^{ - s \sum_{j=1}^N(1/2b_j)}.
    \end{equation}
    
    In the case $\beta = 2$ (general variance complex Wishart matrices)
    the result (\ref{3ee}) has been derived previously \cite{Fo07s}

  \section{The case $\beta = 4$}
  \setcounter{equation}{0}
  The conditional PDF (\ref{Y5}) holds in the case $n \ge N$. If instead $n < N$  the eigenvalues
  $\{y_l \}_{l=1,\dots,N}$ of $\tilde{X} \tilde{X}^\dagger$ will all be strictly positive. With the
  eigenvalue PDF of $X X^\dagger$ determined by (\ref{ePDF}), the corresponding secular equation
  reads
  \begin{equation}\label{sec}
  0 = 1 + b \sum_{j=1}^N {q_j \over y_j - \lambda}
  \end{equation}
  where $q_j \mathop{=}\limits^{\rm d} \Gamma[\beta /2,2]$ $(j=1,\dots,N)$ (cf.~\ref{Y5}).
  
  Let the roots of (\ref{sec}) and thus the eigenvalues of $X X^\dagger$ be denoted
  $\{\lambda_i\}_{i=1,\dots,N}$. Then \cite[Cor.~3]{FR02b} gives that the conditional PDF of
  $\{\lambda_i\}_{i=1,\dots,N}$ given $\{y_j\}_{j=1,\dots,N}$ is proportional to
  \begin{equation}\label{sec1}
  e^{- ( \sum_{j=1}^N (\lambda_j - y_j))/2b}
  {\prod_{1 \le j < k \le N} (\lambda_j - x_k) \over
  \prod_{1 \le j < k \le N} (y_j - y_k) ^{\beta - 1}}
  \prod_{i=1}^N \prod_{j=1}^N | \lambda_i - y_j|^{\beta/2 - 1},
  \end{equation}
subject to the interlacing (\ref{xy}). Let us suppose now that $\{y_j\}_{j=1,\dots,N}$ has
PDF proportional to
\begin{equation}\label{sec2}
\prod_{j=1}^N e^{-y_j/2} \prod_{1 \le j < k \le N} (y_j - y_k)^\beta.
\end{equation}
This is realized by the eigenvalue PDF of the $N \times N$ matrix $B_\beta^T B_\beta$, with
$B_\beta^T$ given by (\ref{Bd}) in the case that $b=1$ and
\begin{equation}\label{su1}
n = N - 1 + 2/\beta.
\end{equation}
Multiplying together (\ref{sec1}) and (\ref{sec2}) shows that the joint PDF of
$\{x_i\}_{i=1,\dots,N}$ and $\{y_j\}_{1,\dots,N}$ is, up to normalization, given by
 \begin{equation}\label{xy3}
  e^{-\sum_{j=1}^N ( y_j + (\lambda_j - y_j)/b)/2}
  \prod_{1 \le i < j \le N} (\lambda_i - \lambda_j) (y_i - y_j)
  \prod_{i,j=1}^N |\lambda_i - y_j|^{\beta/2 - 1},
  \end{equation}
  subject to the interlacing (\ref{sec2}). In the case $\beta = 4$ this reduces to
  (\ref{xy1}).
  
  The results of \cite{BV10} tell us that for a given $\beta$ the parameter dependent soft
  edge state is independent of the ratio $n/N$, provided both $n,N \to \infty$.
  Thus in studying the state we are free to choose a particular dependence of $n$ on $N$,
  which we take to be (\ref{su1}). In \cite{FR02} the correlations corresponding to (\ref{xy3})
  with $\beta = 4$ have been given in terms of a quaternion determinant (Pfaffian) with explicit entries. The correlations were computed in the so-called parity blind case, when the two species
  implied by (\ref{xy3}) --- the $\lambda$'s and the $y$'s --- are regarded as indistinguishable, and
  the parity aware case when they are not. In the parity blind case the soft edge scaled limit
  was also computed.
  
 We will discuss first not the correlations, but the distribution function for the largest eigenvalue.
 The largest eigenvalue belongs to species $x$, so we can equally as well work with
 parity aware, species $\lambda$ case, or the parity blind case; our approach relies on working with the
 latter.
  In particular, in the special case $b=2$,  $\beta = 4$, we see that the parity blind system implied by (\ref{xy1})   is
  precisely the Laguerre orthogonal ensemble with $2N$ eigenvalues and the 
  weight function $e^{-\lambda/4}$, specified by the
  eigenvalue PDF
  \begin{equation}\label{wl}
  {1 \over C} \prod_{l=1}^{2N} w(\lambda_l) \prod_{1 \le j < k \le N}
  | \lambda_k - \lambda_j |^\beta
  \end{equation}
  with $w(\lambda) = e^{-\lambda/4}$ $(\lambda > 0)$, and $\beta = 1$.
  For the PDF (\ref{wl}) in general, let us denote by $E_{2N,\beta}(0;J;w(x))$ the probability of
  no eigenvalues in the interval $J$. We know from \cite{Fo08} that for the Laguerre orthogonal
  ensemble this probability (which is the distribution function for the largest eigenvalue)
  admits the soft edge scaling limit
  \begin{eqnarray}\label{wk}
&&   \lim_{N \to \infty} E_{2N,1}(0;(16N+ 4 (4N)^{1/3} s, \infty); e^{- \lambda/4})  \nonumber \\
&& \quad =   
 \lim_{N \to \infty} E_{N,1}(0;(4N+2 (2N)^{1/3} s, \infty); e^{- \lambda/2})
 = E_{\beta = 1}^{\rm soft}(0;(s,\infty))
  \end{eqnarray} 
  where, with $q(x)$ the transcendent specified by (\ref{qs}), 
\begin{equation}\label{wk1}
\Big (   E_{\beta = 1}^{\rm soft}(0;(s,\infty)) \Big )^2
= \exp \Big ( - \int_s^\infty (x - s) q^2(x) \, dx -
\int_s^\infty q(x) \, dx \Big )
\end{equation}
is the square of the distribution function for the scaled largest eigenvalue in the
Gaussian orthogonal ensemble (GOE; see \cite[eq.~(9.127)]{Fo10}). Hence with $b=2$ the distribution
of the scaled largest eigenvalue in the spiked real quaternion Wishart ensemble is equal to the
distribution of the scaled largest eigenvalue of the GOE, a fact first deduced in \cite{Wa09} using more complicated workings.

Known results can also be used to specify the distribution function for the scaled largest eigenvalue 
of the $\beta = 4$ parity blind system for general $b$ in the scaling
regime about $b=2$. Specifically, set
\begin{equation}\label{bw}
b = 2 - {2^{1/3} w \over N^{1/3}}, \quad w \in \mathbb R,
\end{equation}
and let us denote by $E_{N,\beta}^{\rm spiked}(0;J;b)$ the probability that there are no eigenvalues
in the interval $J$  for the spiked Wishart $\beta$-ensemble with parameter
$b$ given by (\ref{bw}) and $n$ given by (\ref{su1}). We have just seen that
$$
E_{N,4}^{\rm spiked}(0;(s,\infty);b) \Big |_{w=0} =
E_{2N,4}(0;(s,\infty);e^{-\lambda/4}),
$$
and that this in turn permits the scaling limit (\ref{wk}). For general $w \in \mathbb R$ we read
off from \cite[Th.~7.1]{Ba05a} the scaling limit
$$\lim_{N \to \infty} E_{N,4}^{\rm spiked}(0;(16 N+ 4 (4N)^{1/3} s, \infty);b)
= F^\symmO(x;w).
$$

The distribution function $F^\symmO$ is specified in \cite{BR01a} in terms of a Riemann-Hilbert
problem formulation of the Hasting-Macleod solution $q(s)$ (\ref{qs}) of the Painlev\'e II equation,
or equivalently in terms of one member of a Lax pair for $q(s)$. After a slight rewrite
\cite{BV10}, the latter reads
$$
{\partial \over \partial w}
\begin{pmatrix} f \\ g \end{pmatrix}  =
\begin{pmatrix} q^2 & - w q - q'  \\
- w q + q'  &  w^2 - s - q^2 \end{pmatrix} \begin{pmatrix} f \\ g \end{pmatrix} ,
$$
subject to the initial conditions
\begin{equation}\label{fgE}
f(s,0) = g(s,0) = E(s), \qquad E(s) := \exp \Big ( - \int_s^\infty q(t) \, dt \Big ).
\end{equation}
Introducing too the notation
$$
F(s) = \exp \Big ( - \int_s^\infty (t - s) q^2(t) \, dt \Big )
$$
the result of  \cite{BR01a} is that
\begin{equation}\label{fgF}
F^\symmO(x;w) = {1 \over 2} \Big ( (f +g) E^{-1/2} + (f - g) E^{1/2} \Big ) F^{1/2}.
\end{equation}
Note from (\ref{fgE}), (\ref{fgF}) and (\ref{wk1}) that we have
$$
F^\symmO(x;0) =  E_{\beta = 1}^{\rm soft}(0;(s,\infty)),
$$
as is consistent with (\ref{wk}).

Next we turn our attention to the soft edge correlation functions for the $\beta = 4$ case of
(\ref{xy1}). For the finite system,  in \cite{FR02} these have been given in terms of a quaternion determinant (Pfaffian) with explicit entries for both the parity blind and parity aware cases.
However, these are rather lengthy, so we will focus attention on the simplest of the
correlations, $\rho_{(1)}(x)$, which corresponds to the density. Let us use the superscripts ``a''
and ``b'' to denote aware and blind respectively, and furthermore distinguish the two possible aware species $\{x_i\}$ and $\{y_i\}$ by writing a,x and a,y respectively. In the notation of  \cite{FR02}, for  the finite system we have
\begin{equation}\label{tt}
\rho_{(1)}^{\rm b}(x) = {1 \over 2}  f^{22}(x/2,x/2), \qquad \rho_{(1)}^{\rm b}(x) = \rho_{(1)}^{\rm a,x}(x)
+ \rho_{(1)}^{\rm a,y}(x)
\end{equation}
(the factors of 2 in the first equation are due to the difference in scale of (\ref{xy1}) relative to
\cite[eq.~(1.4)]{FR02}). Now, by construction of (\ref{xy1}), $\{y_i\}$ are distributed according to the Laguerre symplectic ensemble (\ref{sec2}) with weight $e^{-\lambda/2}$, and so are independent of the
parameter $b$. Known results for the scaled soft edge correlations in this ensemble
\cite[eq.(7.117)]{Fo10} give
\begin{eqnarray}
 \rho_{(1)}^{\rm a,y}(X) & := & \lim_{N \to \infty} 4 (4N)^{1/3} \rho_{(1)}^{\rm a,y}(16N + 4 (4N)^{1/3} X) \nonumber \\
 & = & {1 \over 2} K^{\rm soft}(X,X) - {1 \over 4} {\rm Ai}(X) \int_X^\infty {\rm Ai}(t) \, dt,
 \end{eqnarray}
 where 
 \begin{equation}\label{KXY}
  K^{\rm soft}(X,Y) := \int_0^\infty {\rm Ai}(u+X) {\rm Ai}(u + Y) \, du
  \end{equation}
  is the Airy kernel.
  Also, from the first equation in (\ref{tt}) we read off from \cite[eq.~(4.27) with $\alpha = w$]{FR02} that
 \begin{eqnarray}\label{KXYa}
 \rho_{(1)}^{\rm b}(X) & := & \lim_{N \to \infty} 4 (4N)^{1/3} \rho_{(1)}^{\rm a,y}(16N + 4 (4N)^{1/3} X) 
 \nonumber \\
 & = & {1 \over 2} K^{\rm soft}(X,X) - {1 \over 2} \int_{-\infty}^X e^{w(X - t)/2} 
 {\partial \over \partial X} K^{\rm soft}(t,X)  \, dt,  \nonumber \\
 && - {w \over 4}  \int_{-\infty}^X  dt \, e^{w(X - t)/2}  \int_X^\infty du \, {\partial \over
 \partial t}  K^{\rm soft}(u,t).
 \end{eqnarray} 
 
 As a check, we see from (\ref{KXY}) and (\ref{KXYa}) that
 $$
 \rho_{(1)}^{\rm b}(X) \Big |_{w=0} = K^{\rm soft}(X,X) + {1 \over 2} {\rm Ai}(X)
 \int_{-\infty}^X {\rm Ai}(t) \, dt.
 $$
 This is precisely the $\beta = 1$ soft edge scaled density for the GOE (see 
 e.g.~\cite[eq.~(7.147)]{Fo10}) as is consistent with (\ref{wk}).
 
 \medskip
 {\it Note added:} The present work was posted on the arXiv in January 2011,
 and a referee report received in July 2011. I've now acted on this report in June 2013
 upon noticing the work \cite{DEKV13} posted on the arXiv and addressing similar material,
 as well as the works \cite{Ru13} and \cite{DL13}, which indicate to me an interest and
 applicability in this line
 of study (the original referee gave the opinion: `Though this new definition is interesting, there
 is little indication how this can be used.' before rejecting it).

\subsection*{Acknowledgements}
This work was supported by the Australian Research Council. I thank
A.~Bloemendal  for discussions, the MSRI Fall 2010 semester on random matrices for making
this possible,  
and D.~Wang for correspondence. I also thank A.~Bloemendal for comments on the first draft
of this work.
 

\providecommand{\bysame}{\leavevmode\hbox to3em{\hrulefill}\thinspace}
\providecommand{\MR}{\relax\ifhmode\unskip\space\fi MR }
\providecommand{\MRhref}[2]{%
  \href{http://www.ams.org/mathscinet-getitem?mr=#1}{#2}
}
\providecommand{\href}[2]{#2}

 \end{document}